\documentclass[doublecol]{epl2} 
\usepackage{amsmath}

\title{Configurations of polymers attached to probes}
\shorttitle{Configurations of polymers attached to probes}
\author{Roy Bubis,\inst{1} 
Yacov Kantor\inst{1} \and Mehran Kardar\inst{2}}
\shortauthor{Roy Bubis, Yacov Kantor \and Mehran Kardar}
\institute{                    
  \inst{1} Raymond and Beverly Sackler School of Physics and
 Astronomy, Tel Aviv University, Tel Aviv 69978, Israel\\
  \inst{2}Department of Physics, Massachusetts Institute of
 Technology, Cambridge, Massachusetts 02139, USA
}

\abstract{
We study polymers attached to spherical (circular) or paraboloidal
(parabolic) probes in three (two) dimensions. Both self-avoiding and 
random walks are examined numerically. The behavior of a polymer 
of size $R_0$ attached to the tip of a probe with radius of curvature $R$,
differs qualitatively for large and small values of the ratio $s=R_0/R$. 
We demonstrate that the scaled compliance (inverse force constant)
$S/R_0^2$, and scaled mean position of the polymer end-point 
$\langle x_\perp\rangle/R$ can be expressed as a function of $s$. 
Scaled compliance is anisotropic, and quite large in the direction 
parallel to the surface when $R_0\sim R$. The exponent $\gamma$, 
characterizing the number of polymer configurations, crosses over
from a value of $\gamma_1$ -- characteristic of a planar boundary -- at small $s$
to one reflecting the overall shape of the probe at large $s$. For a spherical
probe the crossover is to an unencumbered polymer, while for a parabolic probe
we cannot rule out a new exponent.}

\begin{document}
\maketitle

Recent progress in single molecule manipulation \cite{singmol}
enables direct probing of their properties. Most techniques, from
atomic force microscopy \cite{atomic} and  microneedles \cite{needles},
to optical \cite{optical} and magnetic \cite{magnetic} tweezers, 
have one common feature: the investigated molecule is attached to a 
probe comparable in size or larger than the molecule itself. Accurate
interpretation of the experimental results should thus account for the
molecule-probe interactions. Here, we examine some properties -- end-point 
distribution, force-response, and number of configurations -- of simple
polymers attached to spherical or parabolic probes. 

Many properties of long polymer are independent of their microscopic 
details. For example, the mean squared end-to-end distance
of a polymer in a good solvent increases with the number of
monomers $N$ as $R_0^2\sim a^2 N^{2\nu}$, where $a$ is of order
of a monomer size, and the exponent $\nu$ depends only on the space
dimension $d$ \cite{polymers}. This universal behavior
can be explained by the correspondence between polymers in
the $N\to\infty$ limit, and thermodynamic systems approaching
the critical point of a phase transition \cite{polymers,kardar}, and
is exploited in renormalization group treatments of polymers
\cite{schafer}. Such universality also justifies the 
use of self-avoiding walks (SAWs) on lattices \cite{saw} to model
long polymers. While the majority of applications are three-dimensional
(3D), studies of two-dimensional (2D) models are important both as a 
theoretical test bed, and since the effects of self-avoidance
are stronger in lower dimensions: e.g.,  $\nu=3/4$ in 2D while
$\nu\approx 0.588$ in 3D. (The latter value is closer to $\nu=1/2$
which appears in any $d$ in the absence of self-avoidance.)
While self-avoiding walks provide a simple representation of a polymer,
(non-self-avoiding) random walks (RWs) \cite{rw} may capture some
characteristics of polymers in a dense melt \cite{polymers}.
The total number of configurations  ${\cal N}_N$ of walks 
on a lattice scales as
${\cal N}_N\sim \mu^N N^{\gamma-1}$, where $\mu$ is a model-dependent 
connective constant, while the exponent $\gamma$ is universal and equals
to $43/32$ (1.157) for 2D
(3D) SAWs \cite{gamma}, and is 1 for RWs. The universal value of $\gamma$ can 
be modified by introducing geometrical restrictions \cite{binder} that 
are present on all length scales: E.g., the number of SAWs attached by one 
end to an infinite repulsive wall is \cite{whit} 
$\mathcal{N}_{N,\mathrm{wall}}\sim\mu^NN^{\gamma_1-1}$, where
$\gamma_1=61/64$ (0.679) for 2D \cite{cardy} (3D \cite{gammaone}) SAWs, 
and $\gamma_1=1/2$ for RWs \cite{debell}. Similar variations occur when 
a polymer is excluded from a region shaped as a planar wedge in 2D or 3D  
\cite{wedge,cardy}, or a cone in 3D \cite{cone}. 

\begin{figure}
\onefigure[width=8truecm]{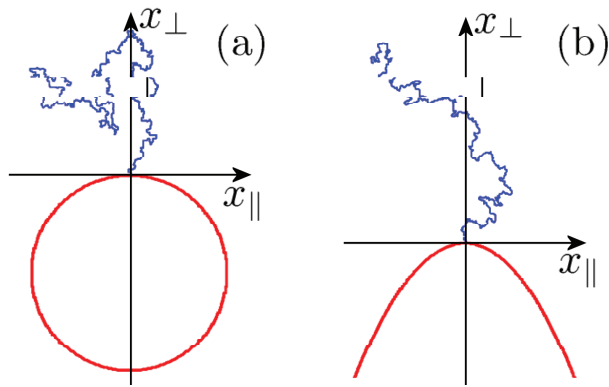}
\caption{(Color online) Self-avoiding walks attached to a repulsive 
(a) circle as in Eq.~\eqref{eq:circle_surface}
or (b) parabola as in Eq.~\eqref{eq:parabola_surface} (sphere or paraboloid in 3D).
The distance $a_0$ between the upper tip of the probe and the origin of the SAW
is equal to a single lattice spacing $a$ and is not visible on this scale.}
\label{fig:models}
\end{figure}

Since real probes usually have rounded end-points, a short polymer (with
$R_0$ smaller than the ``rounding") will behave as if it is in the 
neighborhood of an infinite flat surface. However, a longer polymer will 
be influenced by the overall shape of the probe and exhibit crossover to a 
different asymptotic behavior. In 3D we considered three types of 
shapes: spheres, semi-infinite cylinders and paraboloids. We also
studied 2D versions of these geometries, since in lower dimension
the effects of the probe are stronger. Figure \ref{fig:models} depicts two 
examples of 2D probes whose surfaces are described by
\begin{eqnarray}
\left(x_\perp+a_0+R\right)^2+x_\parallel^2=&R^2,
    \label{eq:circle_surface} \\
x_\perp+Cx_\parallel^2=&-a_0.
    \label{eq:parabola_surface}
\end{eqnarray}
Equation \eqref{eq:circle_surface} defines a circle of radius $R$, and Eq. \eqref{eq:parabola_surface} a parabola with curvature $2C=1/R$ at its
tip. Note that at length scales not
exceeding $R$ the tip of the parabola is similar to the sphere of
the same radius. We model the polymer attached to the probe 
by  a RW or SAW which begins at the origin of 
coordinates, at a small distance $a_0$ away from the nearest point of the
probe surface. The 3D generalization of these surfaces are a sphere and 
a paraboloid. The axis $x_\perp$ is perpendicular to the surface
near the origin, while $x_\parallel$ is parallel to it: In
3D $x_\parallel$ spans the 2D plane parallel to the surface, and $x_\parallel^2$
in Eqs.~\eqref{eq:circle_surface}, \eqref{eq:parabola_surface} should be 
replaced by, say, $x_2^2+x_3^2$.  We may expect that very long polymers will 
see the sphere as a point-like obstacle weakly 
influencing their behavior, while the cylinder will resemble
a semi-infinite excluded line with a larger influence. In either
case no length scale should remain visible to long polymers. Paraboloids and 
parabolas present a slightly more complicated problem: Their width increases
as the square-root of the distance from their tip, thus presenting an 
additional (varying) length scale. The presence of such a sublinearly widening
boundary is thought to be insufficient to modify the asymptotic behavior: 
From the perspective 
of renormalization group, rescaling of all dimensions by a factor of 
$\lambda$ will increase $C$ by only $\lambda^{1/2}$, i.e. make the 
shape approach a semi-infinite line, and possibly behave
as such. Nevertheless, we find that the excluded shape results in
interesting effects \cite{peschel} at finite $N$. 
In this paper we present some results pertaining to circles (spheres) and parabolas 
(paraboloids) in 2D (3D). Additional properties and geometries will
be described elsewhere \cite{bkk}.

Our numerical simulations were performed on a square (cubic) lattice in 
2D (3D) with lattice constant $a$. The probe positions were also
discretized and thus rough on the same scale. (For 
$R\gg a$ we expect discretization effects to be negligible.) We further 
assume that $a_0=a$ does not play a role in the behavior of the polymer. 
Results were obtained using standard MC procedures: the 
samples of SAWs were generated using dimerization \cite{dimer} and 
pivoting \cite{pivot} algorithms, while RWs were either generated 
randomly, or the end-point distributions were found by solving the 
diffusion equation with the probe surface as an absorbing boundary.

\begin{figure}[t]
\onefigure[width=8truecm]{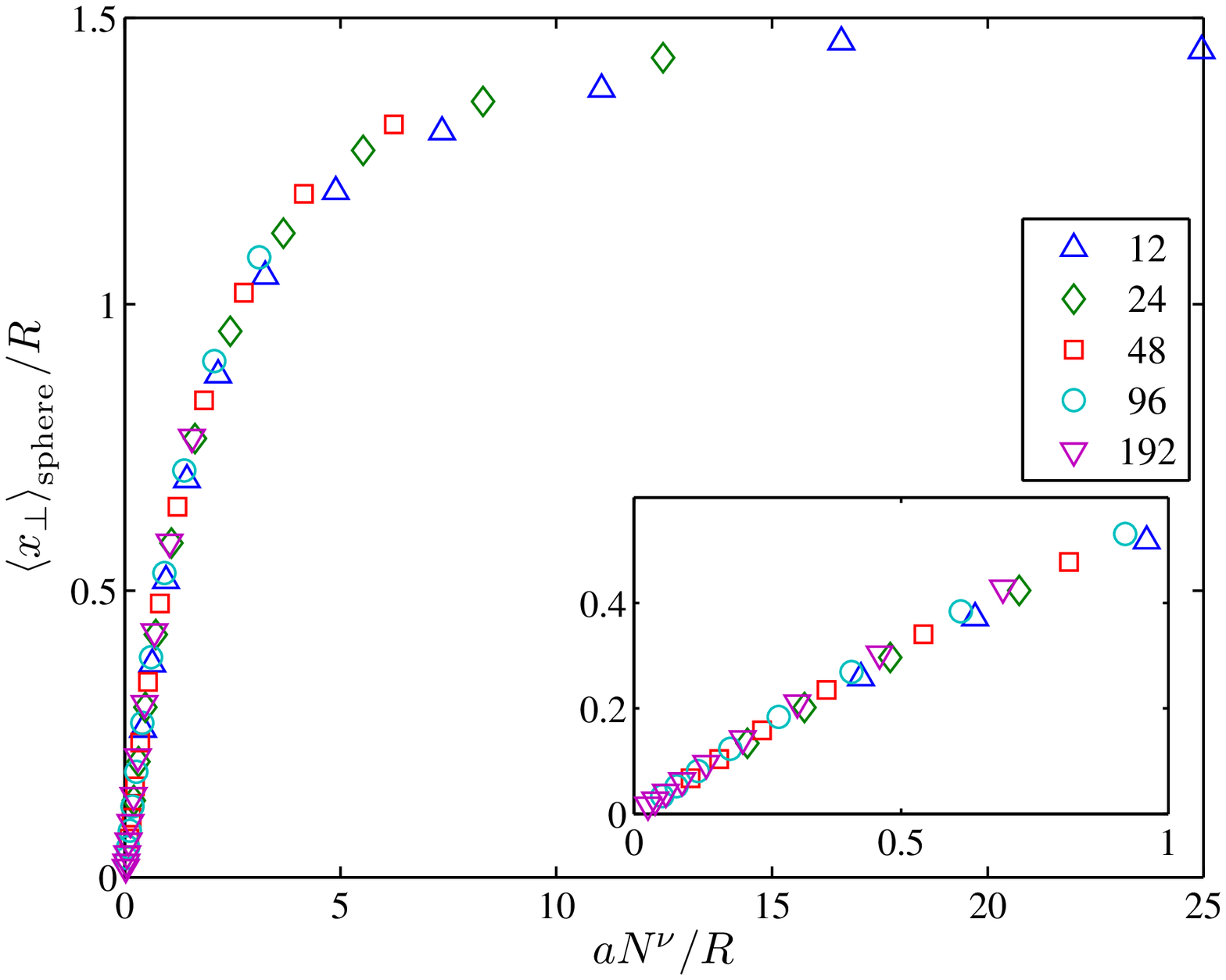}
\onefigure[width=8truecm]{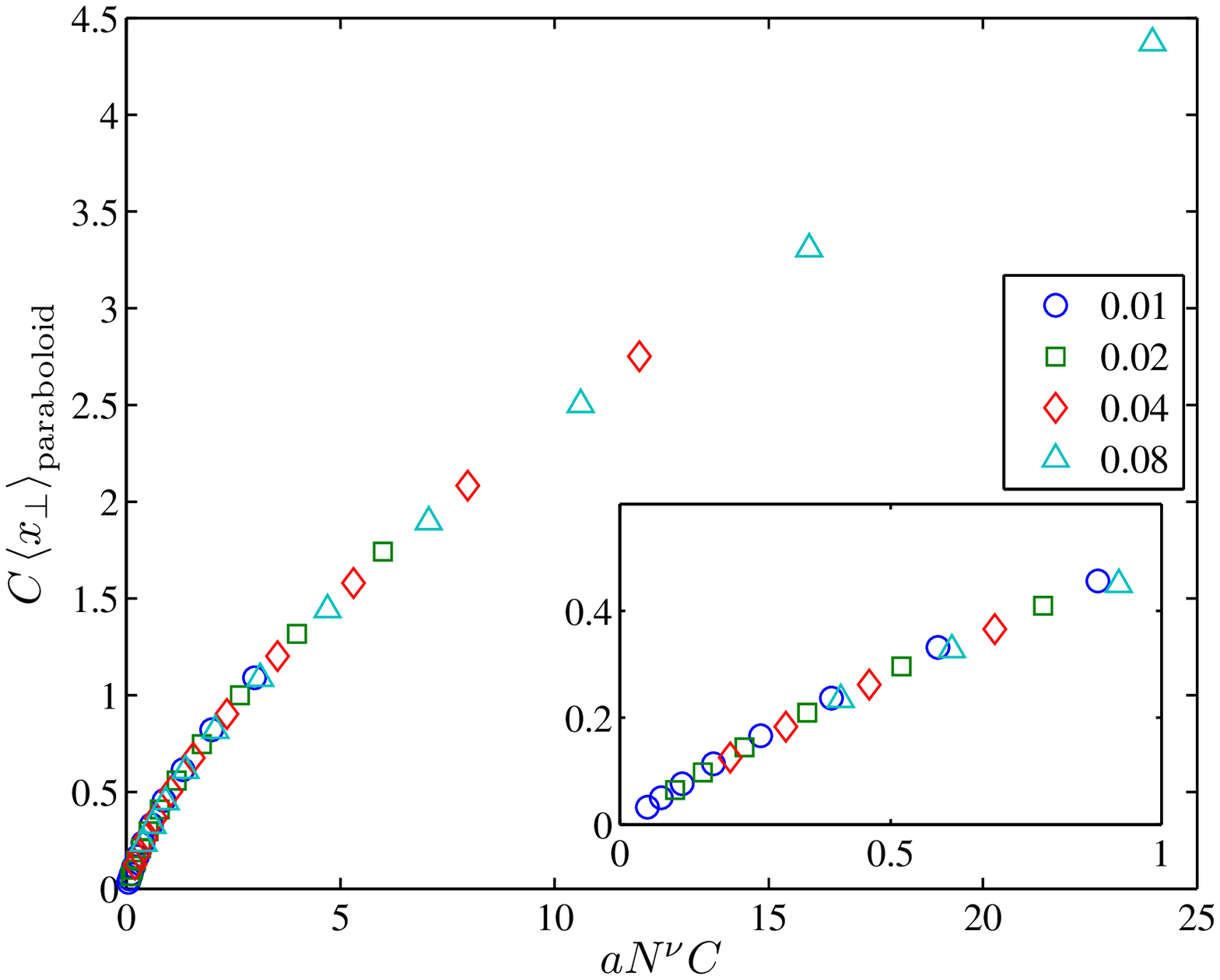}
\caption{(Color online) The scaling function for the mean end-point
position of
3D SAW with $N$ ranging from 16 to 16384 for a sphere (top panel) 
with $R$ from 12 to 192 lattice constants (see legend); and a 
paraboloid (bottom panel) with 
$C$ between 0.01 and 0.08 inverse lattice constants (see legend). The 
insets show the behavior for  small values of (top) $aN^\nu/R$ 
and (bottom) $CaN^{\nu}$.  Error bars are smaller than 
the size of the symbols.}
\label{fig:separation}
\end{figure}

In the absence of an external force the probability distribution of the 
end-to-end vector is the ratio between the number of spatial configurations $\mathcal{N}_N(\vec{r})$ terminating at the position $\vec{r}$ and the 
total number of different configurations $\mathcal{N}_N$ of the walk: $P(\vec{r})=\mathcal{N}_N(\vec{r})/\mathcal{N}_N$. For the
probes depicted in Fig.~\ref{fig:models} the mean position  
$\langle x_\perp\rangle$ of the endpoint of the polymer will be 
along the $x_\perp$ axis. When the lattice constant is significantly shorter
than all other scales of the problem, the probability density 
$P(\vec{r})/a^d$ of the end-point position no longer explicitly
depends on $a$ and $N$, and only depends on the polymer size $R_0$
and the characteristic length $R$ of the probe \cite{polymers}. 
Consequently, $\langle x_\perp\rangle$ which has dimensions 
of length must behave as $R\Phi(R_0/R)$, where $\Phi$ is a dimensionless 
function of the variable $s=R_0/R$. (This argument applies to both
spheres and paraboloids. However, the function $\Phi$ will be different
in each case.)  Figure \ref{fig:separation} depicts $\Phi(s)$ calculated 
for SAWs near a sphere and a paraboloid in 3D. The results for several values 
of $R$ and a large range
of $N$ nicely collapse, confirming our scaling assumption. 
For $s\ll 1$ the polymer does not ``feel" the curvature of the surface, 
and the result should be independent of $R$. This requires that 
$\Phi(s)\sim s$ for small $s$. Indeed, the insets depict the expected 
linear small-$s$ behavior. Moreover, the slopes obtained for the
sphere and parabola coincide, because in this range they are
independent of the shape of the probe. (The factor 2 in the
relation $C=1/2R$ does is not influence these graphs because both axes
include $C$ as a prefactor.)  Similar small-$s$ behavior 
was obtained for SAWs in 2D and for RWs in both 2D and 3D. 

For large $s$, the behavior of $\Phi(s)$ depends on the space dimension,
and the type of walk used to model the polymer, as well as on the type of the 
probe. The behavior near a sphere is the simplest: For SAWs the scaling
function in top Fig.~\ref{fig:separation} approaches a constant, i.e. 
$\langle x_\perp\rangle$ stops increasing. We find a similar behavior
for a RW \cite{bkk}. Both cases are consistent with the fact that 
the mean distance between the origin and the endpoint of a SAW or RW that
starts near an excluded point in 3D is finite \cite{considine}.
The situation is different in 2D: We find \cite{bkk} that the presence 
of the excluded circle causes a logarithmic divergence of $\Phi$ for RWs,  
similar to the behavior in the presence of an excluded point 
\cite{weiss,considine} (since the RW keeps returning to the origin). 
The behavior of 2D SAWs in the presence
of an excluded point is less clear, and arguments for both diverging
$\Phi$ \cite{divslow,divlog} and convergence to a constant \cite{divconst} 
have been advanced. Our results \cite{bkk} were unable to distinguish 
between convergence to a constant and a slower-than-logarithmic divergence.

The behavior of a polymer near a parabola or paraboloid deserves more
careful examination. Even the presence of a semi-infinite excluded line 
has severe effects on the 
mean position of the end-point. In 2D that point moves away from the
starting point as $aN^\nu$ for both RWs and SAWs. Clearly 
$\langle x_\perp\rangle\sim R_0$ (i.e. $\Phi(s)\sim s$) is as far 
away as the end-point is likely to go and therefore, unsurprisingly, we get the 
same result for a parabola \cite{bkk}. The separation of the end-point
in 3D is more striking: For the semi-infinite line the mean
position of the end-point of a RW moves away a distance 
$aN^{1/2}/\ln N$ \cite{considine}. This result is barely distinguishable 
from the maximal conceivable separation $aN^{1/2}$ and
our results for the parabolic probe closely follow this prediction.
More interestingly, the mean position of an end-point of SAWs attached to
the tip of a semi-infinite line in 3D is reported to increase
as $aN^\sigma$ with $\sigma\approx 0.4$ \cite{considine,sigma,sigma_arg}. 
If we assume that this result is also valid for a paraboloid we must have
$\Phi(s)\sim s^\alpha$, with $\alpha=\sigma/\nu$. The high-$s$ end of the
data depicted in the Fig.~\ref{fig:separation}(b) can be fitted
with $\alpha\approx 0.71$ which is close to the value  $0.68$ obtained in the 
simulations of polymers near a semi-infinite line \cite{considine}.

Force-displacement characteristics are quite relevant to experiments. 
The necessary information to obtain such relations is contained
in the details of the end-point probability distribution $P(\vec{r})$.
This distribution plays the role of the Boltzmann weight in this 
ensemble, since the energy is constant and  each configuration has the 
same probability. When a force $\vec{f}$ is 
applied to the end-point of the polymer, the probability distribution 
is shifted by a corresponding Boltzmann weight, such that the mean position 
of the end-point of the walk is obtained by
\begin{equation}\label{eq:rf}
\langle\vec{r}\rangle_{\vec{f}}=
\frac{\int \vec{r}\,P(\vec{r}){\rm e}^{\vec{f}\cdot\vec{r}/k_BT}d^d\vec{r}}
{\int P(\vec{r}){\rm e}^{\vec{f}\cdot\vec{r}/k_BT}d^d\vec{r}}\,.
\end{equation}
The probability $P(\vec{r})$ in the presence of a repulsive probe is not 
isotropic, and therefore the position of the end-point is not necessarily 
directed along the force. However, since the system is symmetric with 
respect to $x_\perp$, by expanding Eq.~\eqref{eq:rf} to the first order in
$\vec{f}\cdot\vec{r}/k_BT$, we find
\begin{equation}
\langle\vec{r}\rangle_{\vec{f}}=(\langle x_\perp\rangle +
S_\perp f_\perp){\bf \hat x}_\perp +S_\parallel f_\parallel{\bf\hat x}_\parallel\,,
\end{equation}
where $\langle\rangle$ without the subscript $\vec{f}$ denotes the 
averages in the {\em absence} of
external force, and the compliances $S_{\perp,\parallel}$ (inverse force 
constants) are given by the variance of the end-point position with zero
force, i.e. $S_\perp={\rm var}(x_\perp)/k_BT$ and 
$S_\parallel={\rm var}(x_\parallel)/k_BT$.
Such linear response is only valid for small forces, i.e. when $f\ll k_BT/R_0$.
This requirement corresponds to force-induced displacements that are
smaller than $R_0$. The large force regime is not considered in this work, 
but may be treatable by using the concept of blobs \cite{polymers}.
Should the experimental setup make it necessary, these results could be
extended to probes attached to both ends of the polymer.

\begin{figure}
\onefigure[width=8cm]{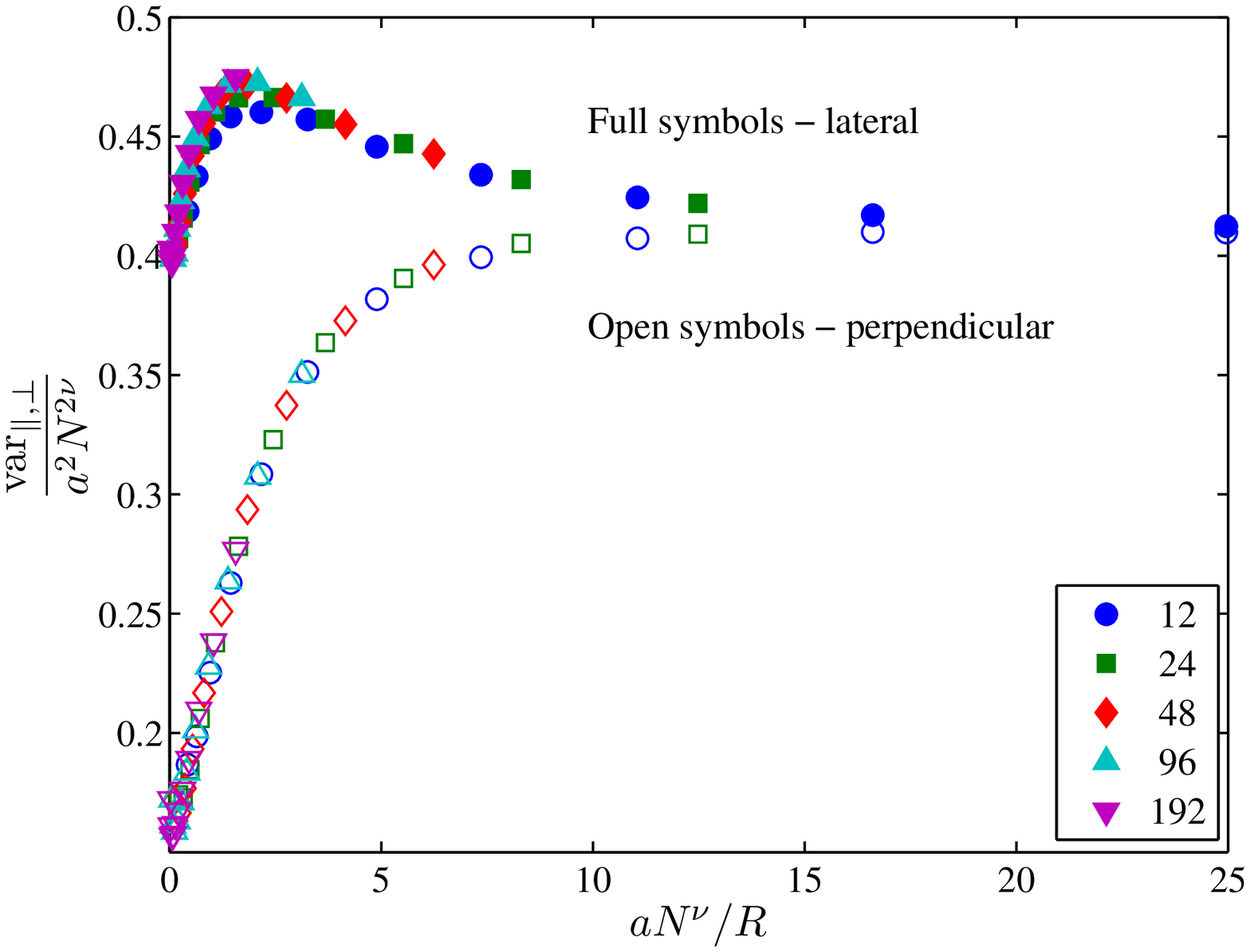}
\onefigure[width=8cm]{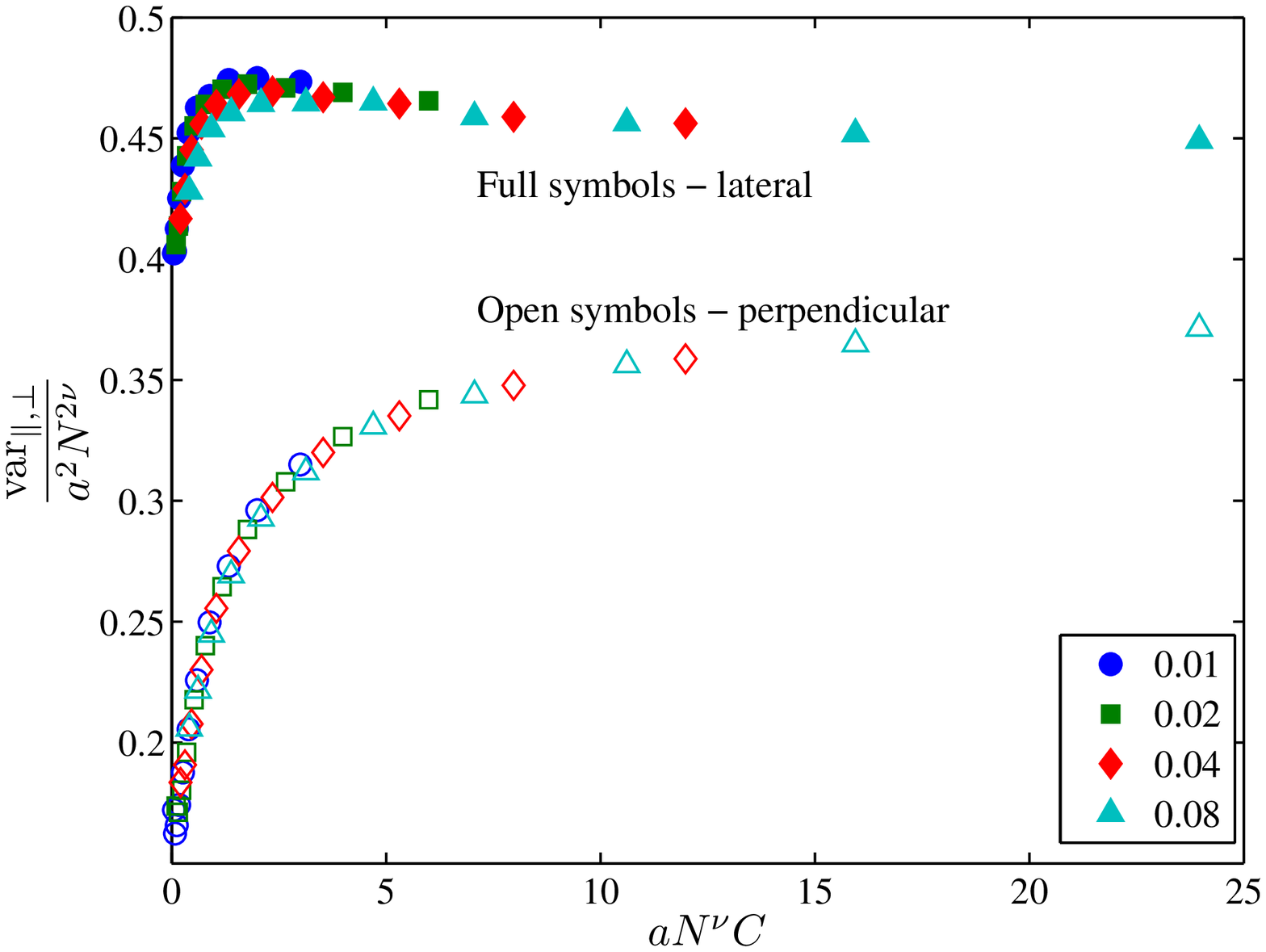}
\caption{(Color online) The scaling function for the linear response to 
force of a three-dimensional SAW near a sphere (top panel) with $R$ 
from 12 to 192 lattice constants (see legend) and  near a paraboloid
(bottom panel) with $C$ between 0.01 and 0.08 inverse lattice constants 
(see legend), for $N$ ranging from 16 to 16384. Open (full) symbols 
indicate response to a perpendicular (lateral) force. Error bars are 
smaller than the size of the symbols.}
\label{fig:var_3D_SAW}
\end{figure}

Since the variance of the position of end-point has dimensions of 
squared length, it can be expressed as $R_0^2 \Phi_{\perp,\parallel}(R_0/R)$. 
Consequently, the ratio between the variance  and $R_0^2$ should only depend 
on $s=R_0/R$, and not separately on $R$ or $N$. The scaling functions
$\Phi_{\perp,\parallel}(s)$ will be different for spheres and paraboloids
and will depend on space dimension. Figure \ref{fig:var_3D_SAW} 
demonstrates nice collapse of data obtained for different values of $R$ 
or $C$ and a large range of $N$. For small $s$ we note that all the curves
approach a constant indicating that, as expected, the compliance is 
proportional to $R_0^2$. There are no differences between the spherical
and paraboloidal probes in this limit since the polymer behaves as if
it is near an infinite plane. We note a significant difference 
between the lateral and perpendicular compliances - the former is
almost thrice the latter, i.e. it is easier to displace the end-point 
of the polymer parallel to the plane. For large values of $s$ the 
differences between the two compliances disappear in the case of a sphere, 
since it no longer influences the polymer. However, quantitative 
differences persist for the paraboloid, since the presence of the probe
is strongly felt for any $R_0$ as already noted in measurements of
the mean position of the end-point.

\begin{figure}
\onefigure[width=8cm]{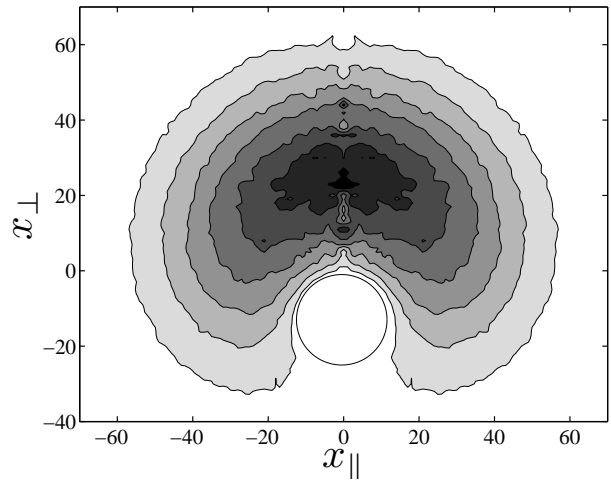}
\caption{
Probability distribution of the endpoint of a 3D SAW with $N=512$ near
a sphere of radius $R=12a$. The contour plot shows a cut through $x_3=0$.
Darker color indicates higher probability density. Contour lines are
equally spaced on linear scale. Raggedness of the lines is a consequence
of discreteness of the lattice.
}
\label{fig:contour}
\end{figure}

Both for spheres and paraboloids, $\Phi$ interpolates  
between the expected small-$s$ and the very-large-$s$ limits. 
Less expected is the non-monotonic behavior of $\Phi_\parallel(s)$: 
the function has a maximum for $s$ of order unity, i.e. when the 
size of the polymer becomes comparable with the typical 
length-scale of the probe. This feature persists for SAWs in 2D and 
is equally pronounced for RWs both in 2D and in 3D\cite{bkk}. 
It should be noted that for any value of $R_0$, the variance of the
position of the end point is of the order of $R_0^2$. That variance 
(and consequently, the compliance constant) remains a monotonically 
increasing function of $R_0$. However, the {\it prefactor} for 
lateral fluctuations becomes somewhat larger when $R_0$ is of the 
same order as the probe. To understand this effect we examine the 
probability density of the end-point of a SAW attached to a sphere. 
Figure \ref{fig:contour} depicts  a cut through the
three dimensional probability for $x_3=0$. This contour plot presents a
case when the size of the polymer is comparable with the radius of 
the sphere. The strong distortion caused by the sphere (the 
probability  density appears to ``hug" the sphere) enhances the 
probabilities at large values of $|x_\parallel|$ at the expense of 
areas close to $x_\parallel=0$, partly explaining the larger variance.

Interesting insights into the statistical mechanics of polymers can
be obtained from a direct study of their free energy, which is proportional
to the logarithm of the partition function $ Z $. The configuration part of
$ Z $ is simply the number of possible states of the polymer in  the presence 
of the probe; its dependence on $N$ exhibits 
interesting crossovers. We will use the sphere to demonstrate the general 
trends. While the exponentially increasing part of the number of states
($\mu^N$) is unaffected by the presence of the probes, the
power law dependence of the number of states on $N$ is modified. We know 
that for 
$R_0\ll R$ the sphere's surface is indistinguishable from
an infinite plane, and therefore 
${\cal N}_{N,{\rm sphere}}\sim\mu^N N^{\gamma_1-1}$, 
as expected for a walk near a wall 
\cite{cardy,gammaone,debell}. Consequently,
${\cal N}_{N,{\rm sphere}}/{\cal N}_N\sim N^{\gamma_1-\gamma}$.
On the other hand, when  $R_0\gg R$, the behavior is expected to match 
that of a walk near an excluded point. Since the ratio between 
${\cal N}_N$ and the number of walks with excluded point remains finite, 
we expect ${\cal N}_{N,{\rm sphere}}/{\cal N}_N$ to approach a constant. 
The crossover between two behaviors appears when $R_0\sim R$, and
an appropriate scaling assumption is 
\begin{equation}\label{eq:scaleN}
\frac{{\cal N}_{N,\rm{sphere}}}{{\cal N}_N}= \left(\frac{R}{a}\right)^{\frac{\gamma_1-\gamma}{\nu}} 
\Psi\left(\frac{aN^{\nu}}{R}\right)\,.
\end{equation}
In the limit of short walks we expect 
$\Psi(s\to 0)\sim s^{(\gamma_1-\gamma)/\nu}$, eliminating the 
dependence on $R$, while for $s\to\infty$ the function should approach
a constant. We verified this scaling behavior numerically for both SAWs
and RWs in 3D and 2D, by examining spheres (circles) with several radii
and a wide range of $N$. For small $s$ we obtained power law dependencies
differing only by few percent from the known values. E.g., for 3D SAWs
near a sphere we found $(\gamma_1-\gamma)/\nu\approx-0.81$, while the 
expected value is -0.78 \cite{polymers,gamma,gammaone}. We obtained 
reasonable data collapse for all cases \cite{bkk} confirming the scaling 
form in Eq.~\eqref{eq:scaleN}. (The 2D case of RWs is slightly different 
since the fraction of walks not returning to the vicinity of the 
origin decreases as $1/\ln N$, thus introducing a slight correction.)

For a semi-infinite cylinder or rectangle of width $2W$ one can 
modify Eq.~\eqref{eq:scaleN} to read
\begin{equation}\label{eq:scalecylinder}
\frac{{\cal N}_{N,W}}{\cal{N}_{N}}= 
\left(\frac{W}{a}\right)^{\frac{\gamma_1-\gamma}{\nu}} 
\Psi\left(\frac{aN^{\nu}}{W}\right)\,.
\end{equation}
As in the case is a sphere, for small $s$ we expect 
$\Psi\sim s^{(\gamma_1-\gamma)/\nu}$, eliminating the dependence
on $W$. In the limit of large $N$ we expect the behavior near
a semi-infinite line. The  latter has been extensively studied:
In general, the number of states of a walk attached to the tip of
a semi-infinite line increases as $\mu^N N^{\gamma_{\rm line}-1}$. 
In 3D the presence of the semi-infinite line does not influence the 
number of states, i.e. $\gamma_{\rm line}$ for SAWs coincides with 
$\gamma$ of unrestricted walks \cite{sigma_arg}. 
(For RWs there is a logarithmic correction \cite{considine}.) 
In 2D the presence of the line has a more significant effect and 
$\gamma_{\rm line}=76/64$ for SAWs \cite{cardy,wedge}, and 
$\gamma_{\rm line}=3/4$ for a  RW  \cite{considine}. Consequently,
for a rectangular/cylindrical probe at large $s$ we must assume that
$\Psi\sim s^{(\gamma_{\rm line}-\gamma)/\nu}$, leading to the
large-$N$ dependence 
\begin{equation}\label{eq:largeNcylinder}
\frac{{\cal N}_{N,W}}{\cal{N}_{N}}\approx
\left(\frac{W}{a}\right)^{\frac{\gamma_1-\gamma_{\rm line}}{\nu}} 
N^{\gamma_{\rm line}-\gamma}\,.
\end{equation}
Our numerical results confirmed this type of behavior for RWs and 
SAWs, both for rectangles in 2D and for cylinders in 3D \cite{bkk}.

The number of states of a polymer near a parabola (paraboloid) presents
a more challenging problem. At length scales shorter
than the curvature $R=1/2C$ of the tip we find results indistinguishable
from the behavior near a sphere. However, for large $N$ curvature
is no longer the relevant length scale since the parabola keeps
widening. At every length scale we may think of the parabola (paraboloid)
as bounded from inside by a cylinder of width $W=a$ and from outside
by a cylinder of width $W=\sqrt{aN^\nu/C}$. These two geometries
provide upper and lower bounds on the number of states of the
polymer near a parabola (paraboloid) ${\cal N}_{N,{\rm par}}$. If
${\cal N}_{N,{\rm par}}/{\cal N}_{N}\sim N^{\gamma_{\rm par}-\gamma}$
then $(\gamma_{\rm line}+\gamma_1)/2\le\gamma_{\rm par}\le \gamma_{\rm line}$.
Our numerical studies of ${\cal N}_{N,{\rm par}}$ produced rather low
quality data collapse, and Table~\ref{tab:one} presents our estimates
for the exponent $\gamma_{\rm par}$, for RW/SAW in 2D/3D. The few percent
statistical errors of each of those estimates is probably smaller than
the possible systematic error. All our results were within the exact bound
on the exponent also presented in Table~\ref{tab:one}. (The
bounds were calculated from the values of $\gamma$ that are either 
known exactly or with high numerical accuracy, and therefore the uncertainties
are smaller than the last included digit.) Our range of lengths was 
insufficient to ascertain if these are true power laws, or crossover effects.

\begin{table}
\caption{Exponent $\gamma_{\rm par}$ near a parabolic probe: numerical estimates 
{\em vs.}  lower and upper bounds}
\label{tab:one}
\begin{tabular}{|l||c|c|c|c|c|c|}
\hline
&\multicolumn{3}{|c}{2D parabola}&\multicolumn{3}{|c}{3D paraboloid}\\
\hline\hline
 &num.&lower&upper&num.&lower&upper\\
\cline{2-7}
RW&0.70&0.63&0.75&0.82&0.75&1.00\\
SAW&1.14&1.07&1.19&0.99&0.92&1.16\\
\hline
\end{tabular}
\end{table}

In summary, we have demonstrated the strong influence of repulsive probes
on the properties of a polymer. Our results regarding the force
constants demonstrate significant differences between lateral
and perpendicular responses, and an unexpected non-monotonic dependence
of the coefficient of the former. In single molecule experiments 
the applied forces range between 0.1pN and 1000pN, and the weak 
deformation regime is usually at the low end of this range. Most of 
the measurements involve perpendicular forces, although measurements 
of lateral forces also exist \cite{bulga}. As the experimental techniques 
become more refined our results will become more relevant, making
it worthwhile to examine additional geometries.
Further study is needed to understand the behavior of the number of
states near a parabola (or paraboloid) to distinguish between possible
crossover effects and new exponents.

\begin{acknowledgments}
This work was supported by the Israel Science Foundation  under Grant No.\ 
99/08 (Y.K.) and by the National Science Foundation under Grant No.\ 
DMR-08-03315 (M.K.).
\end{acknowledgments}


\begin{thebibliography}{0}

\bibitem{singmol}
 \Name{Bustamante C., Bryant Z. \and Smith S. B.}
 \REVIEW{Nature}{421}{2003}{423};
 \Name{Kellermayer M. S.}
 \REVIEW{Physiol. Meas.}{26}{2005}{R119};
 \Name{Neuman K., Lionnet T. \and Allemand J.-F.}
 \REVIEW{Ann. Rev. Mater. Res.}{37}{2007}{33};
 \Name{Deniz A. A., Mukhopadhyay S. \and Lemke E. A.}
 \REVIEW{J. R. Soc. Interface}{5}{2005}{15}.

\bibitem{atomic}
 \Name{Fisher T. E., Marszalek P. E., Oberhauser A. F., Carrion-Vazquez M. 
   \and Fernandez J. M.}
 \REVIEW{J. Physiol.}{520}{1999}{5}.
\bibitem{needles}
 \Name{Kishino A. \and Yanagida T.}
 \REVIEW{Nature}{334}{1988}{74}.
\bibitem{optical}
 \Name{Horme\~no S. \and Arias-Gonzalez J. R.}
 \REVIEW{Biol. Cell}{98}{2006}{679}.
\bibitem{magnetic}
 \Name{Gosse C. \and Croquette V.}
 \REVIEW{Biophys. J.}{82}{2002}{3314}.
\bibitem{polymers}
 \Name{de Gennes P.-J.}
 \Book{Scaling Concepts in Polymer Physics}
 \Publ{Cornell University Press, Ithaca, New York} 
 \Year{1979}.
\bibitem{kardar}
 \Name{Kardar M.}
 \Book{Statistical Physics of Fields}
 \Publ{Cambridge University Press}
 \Year{2007}.
\bibitem{schafer}
 \Name{Sch\"afer L.}
 \Book{Excluded Volumer Effects in Polymer Solutions}
 \Publ{Springer, Berlin}
 \Year{1999}.
\bibitem{saw}
 \Name{Li B., Madras N. \and Sokal A. D.}
 \REVIEW{J. Stat. Phys.}{80}{1995}{661}.
\bibitem{rw}
 \Name{Hughes B. D.}
 \Book{Random Walks and Random Environments}
 \Vol{1}
 \Publ{Clarendon Press, Oxford}
 \Year{1995};
 \Name{Rudnick J. \and Gaspari G.}
 \Book{Elements of the Random Walk}
 \Publ{Cambridge University Press}
 \Year{2004}.
\bibitem{gamma}
 \Name{Grassberger P., Sutter P. \and Sch\"afer L.}
 \REVIEW{J. Phys. A: Math. Gen.}{30}{1997}{7039};
 \Name{Caracciolo S., Causo M. S. \and Pelissetto A.}
 \REVIEW{Phys. Rev. E}{57}{1998}{R1215};
 \Name{Nienhuis B.}
 \REVIEW{Phys. Rev. Lett.}{49}{1982}{1062}.
\bibitem{binder}
 \Name{Binder K.}
 in \Book{Phase Transitions and Critical Phenomena}
 \Vol{8}
 \Editor{Domb C. \and Lebowitz J. L.}
 \Publ{Academic Press, London}
 \Year{1983}
 \Page{1}.
\bibitem{whit}
 \Name{Whittington S. G.}
 \REVIEW{J. Chem. Phys.}{63}{1975}{779}.
\bibitem{cardy}
 \Name{Cardy J. L.}
 \REVIEW{Nucl. Phys. B}{240}{1984}{514}.
\bibitem{gammaone}
 \Name{Grassberger P.}
 \REVIEW{J. Phys. A: Math. Gen.}{38}{2005}{323}.
\bibitem{debell}
 \Name{De'Bell K. \and Lookman T.}
 \REVIEW{Rev. Mod. Phys.}{65}{87}{1993}.
\bibitem{wedge}
 \Name{Cardy J. L.}
 \REVIEW{J. Phys. A: Math. Gen.}{16}{1983}{3617};
 \Name{Cardy J. L. \and  Redner S.}
 \REVIEW{\it ibid.}{17}{1984}{L933};
 \Name{Guttmann A. J. \and Torrie G. M.}
 \REVIEW{\it ibid.}{17}{1984}{3539}.

\bibitem{cone}
 \Name{Slutsky M., Zandi R., Kantor Y. \and Kardar M.}
 \REVIEW{Phys. Rev. Lett.}{94}{2005}{198303}.
 
\bibitem{peschel}
 \Name{Peschel I., Turban L. \and Igl\'oi F.}
 \REVIEW{J. Phys. A: Math. Gen.}{24}{1991}{1229}.

\bibitem{bkk}
 \Name{Bubis R., Kantor Y. \and Kardar M.}
 to be published.

\bibitem{dimer}
 \Name{Suzuki K.}
 \REVIEW{Bull. Chem. Soc. Jpn.}{41}{1968}{538};
 \Name{Alexandrowicz Z.}
 \REVIEW{J. Chem. Phys.}{51}{1969}{561}.
\bibitem{pivot}
 \Name{Lal M.}
 \REVIEW{Molec. Phys.}{17}{1969}{57};
 \Name{Madras N. \and Sokal A. D.}
 \REVIEW{J. Stat. Phys.}{50}{1988}{109}.

\bibitem{considine}
 \Name{Considine D. \and Redner S.}
 \REVIEW{J. Phys. A: Math. Gen.}{22}{1989}{1621}.
\bibitem{weiss}
 \Name{Weiss G. H.}
 \REVIEW{J. Math. Phys. }{22}{1981}{562}.
 
\bibitem{divslow}
 \Name{Grassberger P.}
 \REVIEW{Phys. Lett. A}{89}{1982}{381};
 \Name{Meirovitch H.}
 \REVIEW{J. Chem. Phys. }{79}{1983}{502};
 \Name{Lim H. A. \and Meirovitch H.}
 \REVIEW{Phys. Rev. A}{39}{1989}{4176};
 \Name{Burnette D. E. \and Lim H. A.}
 \REVIEW{J. Phys. A: Math. Gen.}{22}{1989}{3059}.
\bibitem{divlog}
 \Name{Redner S. \and Privman V.}
 \REVIEW{J. Phys. A: Math. Gen.}{20}{1987}{L857}.
\bibitem{divconst}
 \Name{Eisenberg E. \and Baram A.}
 \REVIEW{J. Phys. A: Math. Gen.}{36}{2003}{L121}.

\bibitem{sigma}
 \Name{Vanderzande C.}
 \REVIEW{J. Phys. A: Math. Gen.}{23}{1990}{563};
 \Name{Grassberger P.}
 \REVIEW{J. Phys. A: Math. Gen.}{26}{1993}{2769};
 \Name{Caracciolo S.,  Causo M. S. \and Pelissetto A.}
 \REVIEW{J. Phys. A: Math. Gen.}{30}{1997}{4939}.

\bibitem{sigma_arg}
 \Name{Caracciolo S., Ferraro G. \and Pelissetto A.}
 \REVIEW{J. Phys. A: Math. Gen.}{24}{1991}{3625};

\bibitem{bulga}
 \Name{Bulgarevich D. S., Mitsui K. \and Arakawa H.}
 \REVIEW{J. Phys. Conf. Ser.}{61}{2007}{170}.

\end{thebibliography}
\end{document}